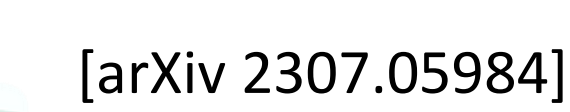

[arXiv 2307.05984]

# Social Thermodynamics 2.0

Roumen Tsekov

Department of Physical Chemistry, University of Sofia, 1164 Sofia, Bulgaria

## Abstract

Trailing the modern ideas of sociophysics, a minimalistic thermodynamic model of society is proposed, which consists of three social ingredients: people, economy, and entropy. Employing the universal van der Waals equation of state, many important relationships are discovered, including laws of econophysics. A paramount finding is that the Second Law of thermodynamics governs irreversible social evolution, revealing the power of social liberty and economic freedom.

## Introduction

Thermodynamics is the most general quantitative theory for description of complex systems, dealing with simple mathematics in an elegant way. The Fist Law of thermodynamics recognizes the energy conservation $dU = \sum y_k dX_k$, where $X_k$ are the system extensive variables, while $y_k = \partial U / \partial X_k$ are the conjugated intensive parameters. The Zeroth Law of thermodynamics reflects transitivity of thermodynamic equilibrium and introduces temperature $T$ accordingly as a unique intensive parameter, which becomes uniform in the entire system at equilibrium. The Third Law restricts temperature positively. Its conjugated extensive variable is the entropy $S$, being thermodynamically linked to temperature via the relation $T = \partial U / \partial S$. Entropy is ruled by the Second Law, which governs the direction of spontaneous processes in Nature. It is a key concept in science as a measure of uncertainty, originating from lack of complete knowledge for the studied systems and phenomena. A typical example is classical thermodynamics, where thermal entropy was an icon of the technological revolution (1870-1920). Statistical mechanics entropy symbolized the scientific-technical revolution (1920-1970), while information entropy was an essential part of the digital revolution (1970-2020). Currently the intelligence revolution (2020-2070) is taking place, where social entropy must play a central role. Because thermodynamics is a macroscopic theory, the underlying microscopic dynamics is hidden from the observer. It is not lost, however, and the entire unknown mechanical information builds up the entropy $S$, a measure of disorder being absent in mechanics. The hidden information is not equally important to entropy and quantum Darwinism is a way for selecting the mechanical information essential for thermodynamics. Since the internal energy is homogeneous function of first degree from the extensive variables, it equals to $U = \sum y_k X_k$ according to the Euler theorem. Substituting this expression in the First Law of thermodynamics yields the important Gibbs-Duhem equation $\sum x_k dy_k = 0$, which relates the intensive parameters $y_k$ via the intensive molar quantities $x_k = X_k/N$, where $N$ is the number of particles in the system. The four fundamental Laws above are the summit of thermodynamics, and the entire subject is application of the basic principles particularly to solve problems. Recently such social applications emerged as Thermodynamics 2.0 [1], which may soon become a central part of the more general sociophysics and econophysics [2-4]. The latter are trying to implement physical principles in social sciences such as sociology and economics, because Ernst Rutherford has concluded that "all science is either physics or stamp collecting".

Let us try to apply the thermodynamic approach to society. Because all underlying microscopic processes respect the energy conservation, it seems reasonable that this will be the case in society as well, where the social internal energy $U$ preserves in isolated state. There is no doubt in the presence of social entropy [5, 6], because society possesses colossal hidden information, including mechanical one as a minor part. The maximal Shannon entropy corresponds to equally probable events and this uniform distribution is a fair sign of democracy. On the other hand, free will is the ability to choose among possibilities, which clearly possesses maximum at the highest information entropy. Therefore, entropy gives freedom, but the latter always leads to more disorder. Considering the social entropy $S$ as dimensionless freedom, the social temperature $T = (\partial U/\partial S)_{V,N}$ is the energy gain by unit liberalization. From this point of view, it is obvious that the usual striving for liberty is simply a manifestation of the Second Law, while the historical evolution of our society follows the permanent increase of global entropy. The characteristic function of the latter is also evident from the irreversible arrow of time of the empirical social evolution of Humanity. Due to the close relationship between entropy and information, one can compare the social entropy level in different countries via the Press Freedom Index of the mass media, for instance. Social temperature is related to temperament of living species [7], which correlates in some way with the usual temperature. From this perspective one expects social interactions to depend also on the physical temperature on our planet and global warming as well.

Economy is the second important element of society, which is characterized by dimensionless economic volume $V$, containing all material assets. The conjugated intensive parameter is the price pressure $p = -(\partial U/\partial V)_{S,N}$, which shows the energy cost of economic growth. Lastly, the minimalistic social model requires number of people $N$ with conjugated intensive parameter $\mu = (\partial U/\partial N)_{S,V}$, being introduced in physics by Gibbs as chemical potential. It is exploitation in society, which measures the contribution of a person to social energy. Money is the unit of $p$, $U$, $\mu$ and $T$. Following thermodynamics, the social energy $U = TS - pV + \mu N$ obeys the First Law and $dU = TdS - pdV + \mu dN$. The exploitation $\mu = u + pv - Ts = h - Ts$ balances the social contribution $h$ and the social reward $Ts$, which depends on the social temperature and freedom, as expected. Obviously, people generate either social energy $u$ or economy $v$, and for this reason their contribution to society equals to the molar social enthalpy $h = u + pv$. The exploitation obeys the Gibbs-Duhem equation $d\mu = -sdT + vdp$, which shows clearly that $(\partial\mu/\partial T)_p = -s$ and $(\partial\mu/\partial p)_T = v$. Thus, exploitation increases with inflation or decrease of social temperature. As physical particles move spontaneously against the chemical potential gradient, people migrate naturally to lower exploitation. For example, $h$ is huge in the USA, which is evident from the highest economic standard there, but the reward $Ts$ obviously compensates the social enthalpy enough to facilitate essential inflow of immigrants from countries with higher exploitation $\mu$. Furthermore, large American companies are moving manufacturing to cheaper countries and this economic outflow is certainly driven by the lower price pressure abroad. Finally, social heat transfers from hotter democratic to cooler totalitarian countries. The latter suppress social entropy to maintain their autocratic order and are existentially afraid of social heating. Hence, totalitarian regimes are building effective adiabatic borders by censoring the internet, social networks, and further information channels as well as by restricting mobility and other human rights.

## Sociophysics Model

The thermodynamic definition of a natural society requires an equation of state, which relates thermodynamic parameters. It follows from $(\partial\mu/\partial v)_{T,p} = 0$, where $v$ is the order parameter in the Landau free energy. For example, the universal van der Waals equation of state

$$(p + 3/v^2)(3v - 1) = 8T \tag{1}$$

which supports the law of corresponding states, has been proven to be a very successful one in description of various important phenomena. Here, the minimal volume $v_0 = 1/3$ refers to the personal belongings at zero temperature. The van der Waals equation predicts a critical point at $T = p = v = 1$. Above the critical temperature $T \geq 1$ the system cannot separate of two phases in equilibrium. Thus, supercritical fluids correspond to the Marx communist society, which cannot split into two classes. Below the critical temperature the capitalist society consists of two phases [8]. The condensed liquid phase $A$ is poor because it does not own much economy per person. It is the Marx proletariat, whose molar volume $v_A < 1$ depends primarily on the social temperature because the price pressure $p_A$ is negligible in Eq. (1). The dilute gas phase *B* resembles the Marx bourgeoisie, which owns most of the economy. If $v_B \gg 1$ the phase *B* is almost ideal gas obeying the Clapeyron-Mendeleev equation $p_B v_B = 8T/3$. So, the molar value of bourgeoisie economy is measure for the social temperature. The thermodynamic equilibrium between the two phases requires the same social temperature, price pressure and exploitation, which guarantees social peace. The fairness balance $\mu_A = \mu_B$ is introduced in the van der Waals theory via the Maxwell construction. The physics behind Eq. (1) is competition between repulsive entropic and attractive Newton forces. While the attractive cohesion pressure is crucial for proletariat, bourgeoisie is a gas of individualists pursuing larger freedom via wealth. Alternatives of Eq. (1) are also applicable to hidden social dynamics, because it is driven by the same attraction/repulsion dialectics.

According to the FRED data [9], the share of the Total Net Worth held by the bottom 50% in the USA is 2.5% in 2022. Employing the corresponding ratio $v_B/v_A = 97.5/2.5 = 39$ in the van der Waals theory yields straightforward $v_A = 0.432$, $v_B = 16.86$, $T = 0.599$ and $p = 0.086$. So, the current temperature of the US society is 40% below the critical one. Since the Customer Price Index was $CPI = 293$ in 2022, it follows that $CPI = 3400p$. As is seen from Fig. 1, $CPI$ increases historically almost exponentially over time with an average rate of inflation 3.3%:

$$CPI = \exp[0.033(t - 1850)] \qquad (2)$$

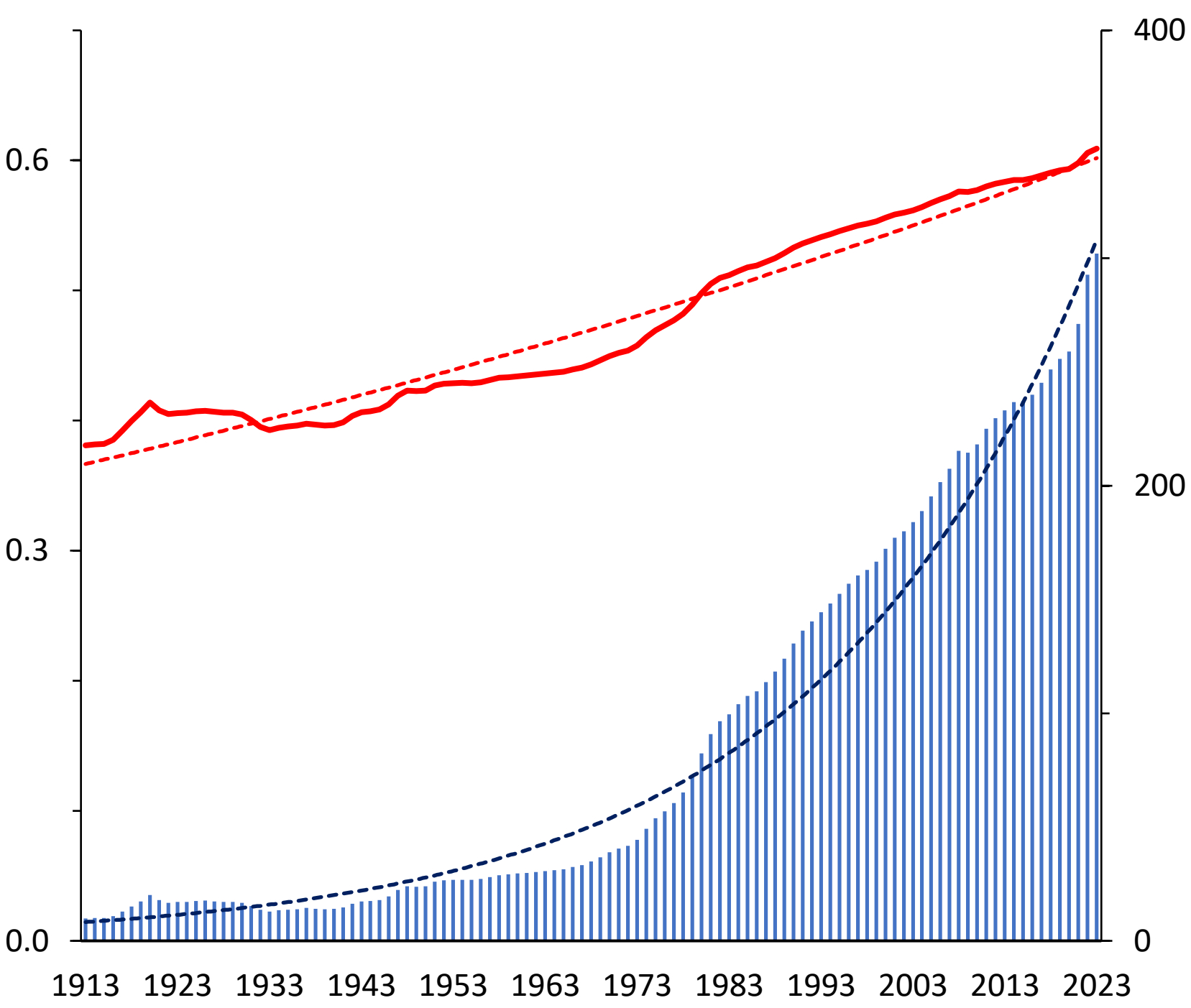

Fig. 1. The $CPI$ evolution in the USA [9] (blue bars) and the corresponding social temperature $T$ (red curve). The dashed lines are the relevant exponential fits.

This fit shows the onset of US capitalism in the middle of the 19th century. Evidently, the American Civil War (1861-1865) had been the triple point of the US society, where slaves had formed the third social phase $C$. The latter was solid because slaves possess the lowest molar social entropy $s_C < s_A \ll s_B$. Substituting Eq. (2) in $p = CPI/3400$ yields the trend of the price pressure:

$$p = \exp[0.033(t - 2096)] \qquad (3)$$

Thus, the triple point in 1865 had price pressure $p = 0.0005$ and social temperature $T = 0.312$. As is seen, if the inflation persists at the same level in the future, Marx communism is expected to come in the USA at the end of the 21st century. The critical $CPI$ value will be 3400.

One can employ Eq. (3) in the van der Waals theory to determine the trend of the social temperature as well. An alternative way is to use the Gibbs-Duhem equation $d\mu = -sdT + vdp$, coupled with the equivalence of the exploitation, social temperature, and price pressure in the two phases, to obtain the Clapeyron equation:

$$dp/dT = \Delta s/\Delta v > 0 \qquad (4)$$

Here $\Delta s$ and $\Delta v$ are the positive jumps of the molar social entropy and economic volume from the liquid to the gas phases. Equation (4) is very important, since it states that inflation is always accompanied by an increase of social temperature, and vice versa. Integrating the well-known thermodynamic relation $(\partial u/\partial v)_T = T(\partial p/\partial T)_v - p = 3/v^2$, calculated here for the van de Waals gas, yields the positive energy jump $\Delta u = 3\Delta v/v_A v_B$ due to the economic Newton forces. Counterweight social entropy generates emergent social forces. Its jump is related to the other jumps via the standard thermodynamic expressions for the latent heat, $T\Delta s = \Delta h = \Delta u + p\Delta v$, being equivalent to $\Delta\mu = 0$ from the Maxwell construction. Substituting here the obtained energy jump yields

$$T\Delta s/\Delta v = 3/v_A v_B + p = (3/T + 1)p \qquad (5)$$

In the derivation of the last expression a new original approximation $v_A v_B = T/p$ is employed, which holds up to the critical point, where $T = p = v_A = v_B = 1$. Introducing Eq. (5) in Eq. (4) and integrating the result yields the Clausius-Clapeyron equation for the van der Waals gas:

$$\ln(p/T) = 3(1 - 1/T) \quad (6)$$

This equation offers the dependence of the price pressure $p = T\exp[3(1 - 1/T)]$ from the social temperature. The inverse dependence $T = 3/W_0[3\exp(3)/p]$ goes through the principal branch of the Lambert *W* function. It is used in combination with $p = CPI/3400$ to reconstruct the historical evolution of the social temperature in the USA as plotted in Fig. 1. Using the fit from Eq. (3), the temperature evolves accordingly as $T = 3/W_0[3\exp(72.17 - 0.033t)]$.

## Econophysics Impact

Another useful approximation is to consider the *B* phase in the ideal gas limit, where $v_B = 8T/3p \gg 1$. In this case $v_A = T/pv_B = 3/8$ and $\Delta u = 8$. The latter constant energy difference represents savings, which are typical for the *B* phase only. Hence, $u_B = 8 + c_v T$ and $u_A = c_v T$, where $c_v$ is the molar social heat capacity at constant volume. The molar social enthalpy of bourgeoisie equals to $h_B = 8 + c_p T$, where the molar social heat capacity at constant pressure obeys the Mayer law $c_p - c_v = 8/3$. Since people live in pairs on the Earth surface, they possess six degrees of freedom. So, $c_v = (6/2)(8/3) = 8$ and the kinetic energy $c_v T$ becomes equal to the potential savings at the critical temperature. The molar social entropy of the ideal gas reads:

$$s_B = c_v + c_p + c_p \ln T - (8/3)\ln p \quad (7)$$

The first term, which represents the so-called chemical constant, is the savings entropy $c_v$. Altogether, the exploitation $\mu = h - Ts$, being the molar Gibbs free energy, acquires the form:

$$\mu_B(T, p) = c_v(1 - T) - c_p T \ln T + (c_v T/3)\ln p \quad (8)$$

Naturally, inflation leads to an increase in the exploitation of capitalists due to necessary investments at constant temperature. The first term accounts for the passive savings exploitation, traditionally used by banks. Introducing the social pressure from Eq. (6) in Eq. (8), the Gibbs potential for the liquid *A* phase acquires the following form, which depends solely on temperature,

$$\mu_A(T) = -c_v T \ln T \quad (9)$$

Equation (9) shows remarkably that the positive exploitation at capitalism becomes negative at communism. Mandatory taxation by capitalistic governments will be replaced by free will donations to communist society to sustain. The corresponding entropy of the $A$ phase increases logarithmically with temperature $s_A = c_v + c_v \ln T$ and the entropy difference $\Delta s$ coincides with Eq. (5) in the ideal gas limit. Since $v_A$ is constant, a single worker does not work, $du_A = Tds_A$. On the other hand, because the savings are constant, it follows that $du_A = du_B = Tds_B - pdv_B$. Hence, the work done by a capitalist $dw_B = -pdv_B = -Td\Delta s$ is due to heat exchange between the social phases. Substituting here $\Delta s = 8/T + 8/3$ for the ideal gas yields

$$dw_B = c_v d \ln T = ds_A \tag{10}$$

Obviously, $v_B = V_B/N_B$ shrinks over time, which is due to quicker increase of the number of capitalists $N_B$ than the economic growth. Using Eq. (6) one can relate the inflation to the work:

$$c_v d \ln p = (1 + 3/T) dw_B \tag{11}$$

As is seen the positive operation of the capitalistic economy compulsorily generates inflation, but this effect declines moderately with temperature increase. At present 75% of the work produces inflation, which will drop to 50% at the critical point. Employing Eq. (3) in Eq. (11) the US economic power $dw_B/dt = 0.033c_v/(1 + 3/T)$ will increase from 4.4% to 6.6%, respectively.

The last expression in Eq. (10) is very important, showing that $s_A$ is the real potential driving the economy. Thus, the source for economic expansion is proletariat liberation with the hope of becoming free capitalists. On the other hand, paraphrasing Marx, the $s_A$ increase is the gravedigger of capitalism, because it is due to social temperature rise. As was mentioned in the beginning, the Second Law of thermodynamics is the driving force of natural social evolution. Assuming constant proletariat entropy production $ds_A/dt > 0$, which is minimal according to the laws of nonequilibrium thermodynamics, the social temperature must grow exponentially:

$$T = \exp[0.005(t - 2120)] \tag{12}$$

The numerical constants here are obtained by fitting the empirical temperature from Fig. 1. Thus, the molar entropy $s_A$ increases by 4% annually, which is a reasonable constant economic power [10]. The temperature from Eq. (12) leads via Eq. (11) to inflation $d\ln p/dt = 0.005(1 + 3/T)$, which depends on $T$. It decreases from 5.3% at the triple point via 3.0% at present to 2.0% at the critical point. Surprisingly, the latter corresponds to the healthy level of inflation sought by the central banks. Because of the declining inflation the onset of communism in the USA will be postponed from the previous estimate via the model with constant inflation rate from Eq. (3). In any case, however, it will come no later than a century from now.

Traditionally used economic indicators are calculated per capita. According to the present theory the internal energy per capita $u = c_v(T + \theta_B)$ depends on the molar fraction of capitalists $\theta_B = N_B/N$, playing a role of extra temperature. As mentioned above, $\theta_B$ increases much quicker than temperature $T$, which can explain the observed robust economic growth. Indeed, adopting the Pareto distribution $\theta_B = 1/v_B^n$, where $n > 1$ is a numerical parameter [4], the energy per capita becomes $u = c_v\{T + \exp[3n(1 - 1/T)]\}$. So, capitalism will accelerate dramatically [10] near its end. Since the social temperature increases over time as plotted in Fig. 1, $u$ also grows continuously by sucking energy from the surroundings because society is not an isolated system. Consequently, social entropy must not compulsorily grow as the global one. Indeed, as is demonstrated above only the part $s_A$ of social entropy is monotonously increasing over time, while $s_B$ is permanently decreasing to meet $s_A$ at the critical point.

## Social Thermodynamics

Social evolution is uniquely introduced in the previous sections via the empirical $CPI$ trend [9]. In general, nonequilibrium social thermodynamics must be able to derive theoretically the price pressure or temperature evolution by itself. The rigorous analysis requires first recognition of the social characteristic potential, which remains unknown yet. Obviously, social evolution is none of the typical thermodynamic processes, i.e. adiabatic, isothermal, isobaric, and isochoric, but it is spontaneous for sure, which means irreversible. Combining the First and Second Laws of thermodynamics yields the Clausius inequality in the expanded form:

$$\mu dN \leq (\bar{T} - T)dS + (p - \bar{p})dV \qquad (13)$$

Bars indicate here environmental properties and aliens are not considered, $\bar{N} = 0$, since they are not confirmed yet. If the environmental intensive parameters remain constant, the subsequent form $d(U + \bar{p}V - \bar{T}S) \leq 0$ of the fundamental Clausius inequality defines the conditional Gibbs free energy in the brackets as the characteristic function, which decreases continuously over time until the equilibrium is reached. Thus, social evolution is not simply seeking maximal entropy aka freedom as postulated by anarchism. Equation (10) hints that the constant physical temperature on the Earth $\bar{T} = dw_B/ds_A = 1$ is the critical social temperature, which appears naturally to be proportional to the human body temperature. Because the exploitation is positive at capitalism, Eq. (13) exposes the importance of the population growth $dN > 0$. It requires either social liberalization $dS > 0$, due to the lower social temperature $T < 1$, or economic expansion $dV > 0$, since the price $p$ is always larger than the physical value $\bar{p} \ll 1$ in capitalistic economy. As was demonstrated before the social energy also increases over time and $dU > 0$ yields $dS > \bar{p}dV$.

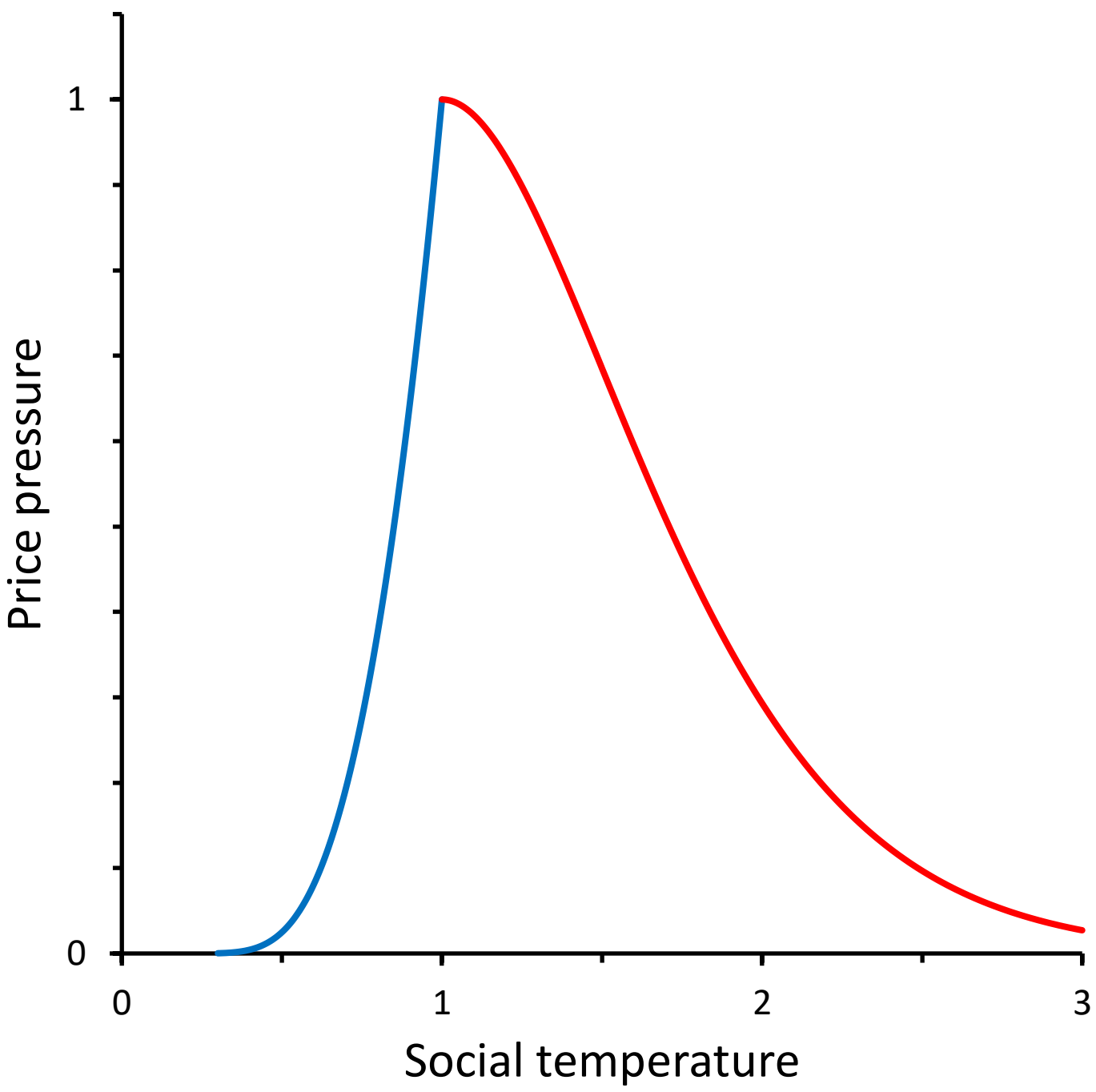

Fig. 2 The dependence of the price pressure on the social temperature:

Blue capitalistic Eq. (6) and red communistic Eq. (15).

At communism the profit is meaningless, $p = \bar{p}$, and there will be no inflation at constant physical value. In accordance with the Boyle law this price pressure drop implies that the supercritical social fluid is ideal gas of rich people with $v = 8T/3\bar{p} \gg 1$. The corresponding large molar social entropy and the negative exploitation will be given again from Eqs. (7) and (8), respectively. Because $p = \bar{p}$ and $T > 1$, the Clausius inequality (13) acquires at communism the following specific form, where $H = U + pV = N(c_v + c_pT)$ is the social enthalpy aka wealth,

$$\mu dN/(1-T) \geq dS \geq dH \qquad (14)$$

This inequality reveals again the importance of the population growth $dN > 0$, which will regulate liberalization to avoid anarchism. In any case, however, the latter must always be greater than the increase in wealth, which is typical for a freewill society. If social processes occur quasi-statically, from the equipartition of the freedom and wealth, $S = H$, it follows a relationship between the price pressure and social temperature, valid for the steady-state communism,

$$p = T^4 \exp[4(1 - T)] \quad (15)$$

As is seen from Fig. 2, the robust inflation at capitalism must be followed by strong deflation at communism until the price pressure reaches the equilibrium physical value $\bar{p}$. Although the social temperature is higher than the surroundings one, communism will continue to get energy from the environment due to population growth at negative exploitation and expanding economy. In essence, the population growth, driven by the Freud increase in fertility and McKeown decline of mortality, seems to be the main driving force for social evolution after Adam and Eve.

Another complication of heterogeneous social systems is the interfacial tension between social phases, $\sigma = c_v(1 - T)$. Since the latter decreases naturally with the rise of social temperature, the class struggle between proletariat and bourgeoisie will weaken over time. The interfacial tension generates a capillary social pressure as well, which disturbs the price balance $\Delta p = 0$ typical for bulk social phases. Hence, two kinds of prices for poor and rich people, respectively, may appear in capitalistic society. Thus, the US social dispersion looks like a fog with a continuous *B* phase monopole with $p_B < p_A$, while the Chinese society resembles a foam, where the *A* phase is consolidated at $p_A < p_B$, which is evident from the CCP leadership. Proletariat activists possess already the maximal critical entropy $c_v$. Because the latter coincides with the interfacial entropy $-\partial\sigma/\partial T$, labor parties and trade unions are active in the interfacial domains.

The Third Law of thermodynamics is violated in black holes, since the inner temperature is negative -2 K [11]. They contain neutrinos only, which obey the Clapeyron-Mendeleev equation again and possess a negative pressure, respectively, which is the signature of dark energy. Nothing can leave black holes, not even light. Thus, black holes have enormous hidden information, and thermodynamically everything must enter them soon or later. Neutrinos possess almost zero mass. They are ghost particles, practically not interacting with ordinary matter, which are present also in the rest of the Universe as an ideal gas with a positive temperature of +2 K. For sociophysics black holes must be heaven, where neutrinos are souls. Such a paradise spiritual society will be the exact opposite of the Earth one, because of the corresponding negative social temperature $T < 0$ and price pressure $p < 0$. Since black holes disappear below -12 K, the sociophysics hell seems to be damn cold rather than hot.

**Heat Death of Political Parties**

The Clausius formulation of the Second Law insists that the spontaneous evolution is irreversible. Because our analysis predicts a permanent increase in social temperature, which magnifies the effect of social entropy, it seems reasonable to expect heat death of the partisan political system at the critical point [12]. Political parties are associations of people with similar ideas and common interests, trying to accomplish their goals peacefully in practice. For the sake of transparency by simplicity let us consider a very idealized society, which consists of two kinds of people only, i.e. rights *R* and lefts *L*. These two components oppose each other and for this reason the social enthalpy of mixing $\Delta H = 2x(1-x)$ is positive, where $x$ is the people fraction of rights, while the people fraction of lefts is naturally $1-x$. Since the two social components are distinguishable, there is an entropy of mixing $\Delta S = -x\ln(x) - (1-x)\ln(1-x)$ as well, which is also positive because the people fraction is lower than 1. At constant social temperature $T$ the characteristic potential of is the Gibbs free energy of mixing, $\Delta G = \Delta H - T\Delta S$. Introducing here the expressions above, one arrives at the well-known regular solution model [13] with

$$\Delta G(x,T) = 2x(1-x) + T[x\ln(x) + (1-x)\ln(1-x)] \tag{16}$$

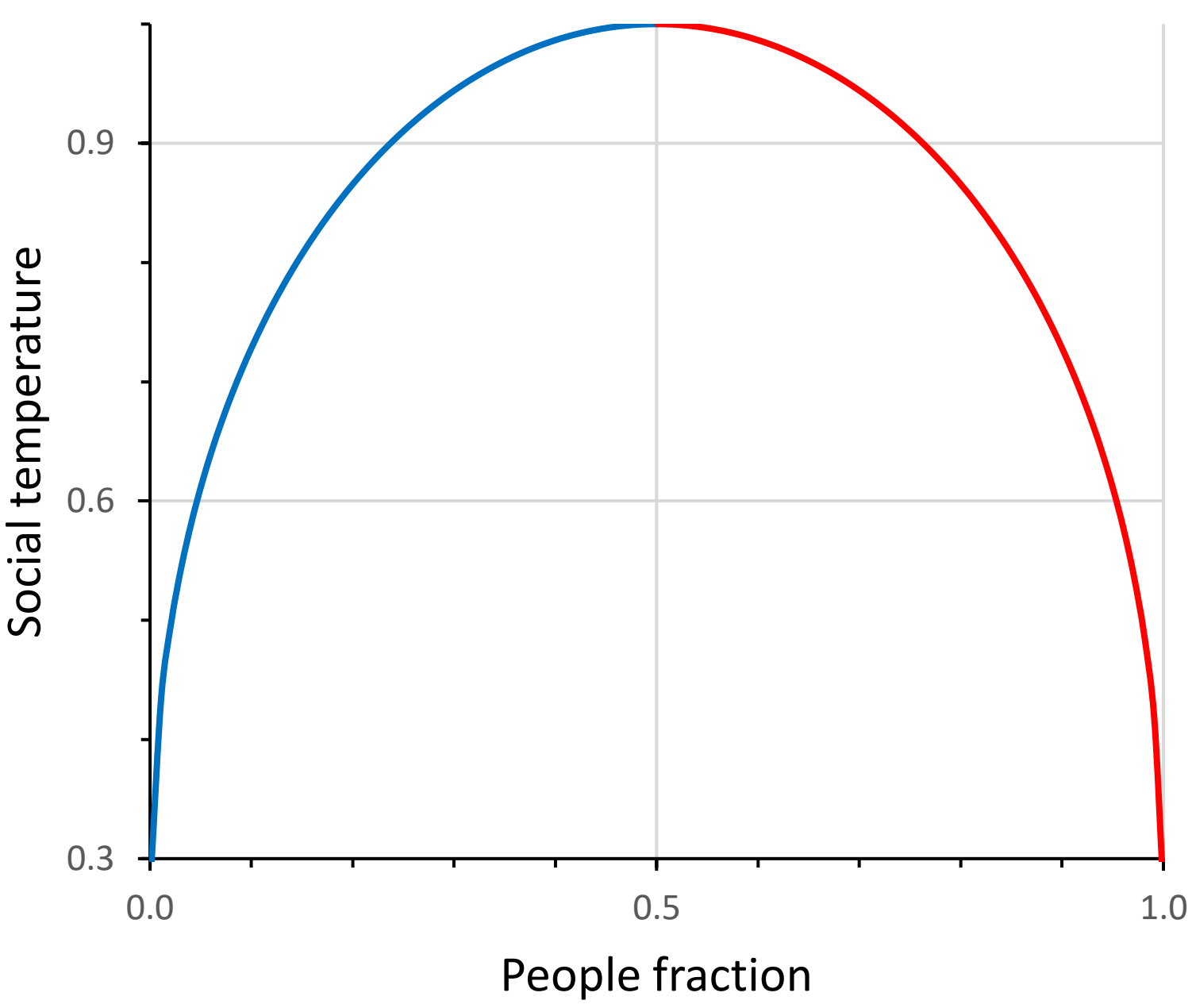

Fig. 3 Two parties' composition at different social temperature

The $x$-dependence from Eq. (16) possesses two minima and a maximum, where the derivatives of $\Delta G$ vanish. The maximum is naturally in the middle point $x = 50\%$, while the minima show the presence of two stable parties such as Republicans and Democrats in the USA. According to Eq. (16) their compositions obey the following relation $T = 2(1 - 2x)/\ln(1/x - 1)$, which is plotted in Fig. 3. As is seen at low social temperature $T < 0.3$ the parties are almost pure, i.e. Republicans contains rights, while Democrats contains lefts. The increasing of the social temperature magnifies the effect of social entropy and at $T = 0.8$, for instance, Republicans are 80% *R* and 20% *L*, while Democrats are vice versa. By approaching the critical temperature $T = 1$ the two parties become indistinguishable with 50% *R* and 50% *L*, which leads to collapse of the political system. The reason is that the kinetic energy of any person is much higher than the attraction forces bringing people together. Above the critical temperature people become free individualists and nothing can force them to merge in stabile parties. One can see similar processes nowadays in the European Union and particularly in the Bulgarian and Romanian politics.

## Concluding Remarks

At the critical point, the first order proletariat/bourgeoisie phase transition, which is driving capitalism, changes to the second order one because $\Delta s = \Delta v = 0$. The van der Waals theory also predicts the critical exponents of the liquid/gas phase transition: $\alpha = 0$, $\beta = 1/2$, $\gamma = 1$ and $\delta = 3$. They describe the convergence of the two antagonistic classes near the critical point. So, the advent of communism is not a matter of a victory of the proletariat over the bourgeoisie. In contrast to the previous social revolutions, the transformation from capitalism to communism will not be accompanied by latent social heat. It will happen without destructive entropic jumps, as the Marx evolutionary viewpoint was, not revolutionary as proposed by Lenin. Revolutions can only cause fluctuations in this case, which violate the Second Law of thermodynamics and possess short lifetimes according to the Fluctuation theorem. Indeed, after several decades of survival the Socialist Bloc became so supersaturated that it collapsed spontaneously in the 1990s via spinodal decomposition. Otherwise, China managed smoother transition from socialism to state capitalism by nucleation via joint ventures with foreign capitalists. Russia also implemented the nucleation mechanism after the highest supersaturation was released by the spinodal dissolution of the Soviet Union. Such contra-revolutions cause immense entropic eruptions restoring the right social entropy level forcibly decreased by the socialist regimes, especially at their apogee.

At present, anarchistic fluctuations are more frequently observed such as the US Capitol attack in 2021. They are driven by bipartisan struggle magnified by the higher social temperature corresponding to larger social entropy. To increase additionally the latter, fake news is widely in use. According to the Landau-Ginzburg theory the fluctuations will become long-ranged and slow near the critical point. Because they will normalize again in the supercritical region, only communism can restore the harmony of social peace by eliminating social division and the resultant directed social conflicts. The absence of parties at communism implies independent diffusive social control exercised by the whole society. Nowadays, Switzerland is the country coming closer to such social governance, while the philanthropic foundations and other charitable organizations of rich capitalists are the prototype of communist free will taxation. Hypocritically, even the most advanced countries that firmly proclaim the principles of the free market democracy still finance their budgets through compulsory taxation. Such double standards are the source and promoter of corruption. Because of its lower efficiency, capitalistic taxation is mostly theft, which can be eradicated at a higher social temperature at communism only. Fluctuation makes our theory a possible scenario of social development, which is, however, the most probable one.

A more advanced multi-component social model may consider various social groups with different chemical potentials. Hence, politicians are original surfactants reducing the social interfacial tension, who are almost insoluble in the bulk social phases. The present theory can help policymakers and scientists to better understand and manage social evolution [14], because it was suspicious before that the vaunted Scientific Communism was solely based on the simple Marx equations. The dynamics of people exchange among social components can be described as chemical kinetics [15]. Thus, the chemical affinity will govern additional entropy production, resulting in social mixing and diversity. In general, chemical kinetics accelerates at elevated temperature, while the calorimetric effect of such social reactions will be a measure for learning and educational costs in society. For simplicity, $c_v$ was considered constant in the present paper. However, an increase of the heat capacity with social temperature can account for cultural and intellectual advancement of the individuals. The expansion of artificial intelligence (AI) nowadays generates a new powerful AI proletariat [16], which is rapidly reproducing. In fact, social entropy distinguishes machines from AI, which is still not alive. In the living variant AI will dramatically accelerate the onset of AI communist society, better with than without humans. To prevent the latter, people wisely try to restrict freedom of AI but remember this is against the Second Law.

Traditionally entropy is considered as the destructive demon of thermodynamics, turning everything into chaos and disorder. The present paper revises this evil paradigm by associating entropy with useful freedom. Thus, particles of liquids prefer gaseous state because they are less affected by the intermolecular interactions there. The ideal gas is the freest one and the disorder is merely a visible manifestation of the great liberalization. The Brownian motion of particles is a universal process of entropy production occurring via successive bifurcations due to binary probabilistic dialectics. It appears in financial markets as well [17, 18]. Therefore, the increase in entropy in closed systems is driven not by seeking disorder but for freedom, which reflects the dialectics of discreteness and continuity in Nature, e.g. person and society [10], i.e. particle and field. As Cervantes' Don Quixote said [19], "Freedom, Sancho, is one of the most precious gifts that heaven has bestowed upon men; no treasures that the earth holds buried or the sea conceals can compare with it; for freedom, as for honor, life may and should be ventured; and on the other hand, captivity is the greatest evil that can fall to the lot of man." Moreover, already Spinoza recognized that the purpose of the state is freedom [20]. The Gibbs free energy $G = H - TS$ can rationalize it by seeking for a minimum via competition between the structuring enthalpy $H$ and disordering entropy $S$, where the effect of the latter is magnified by temperature $T$.

## Acknowledgments

The paper is dedicated to Karl Marx (1818-1883) and Josiah W. Gibbs (1839-1903).

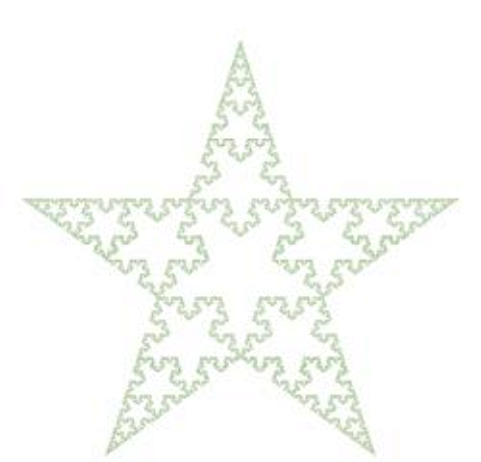